\documentclass[aps,prl,superscriptaddress,twocolumn,floatfix,preprintnumbers]{revtex4}
 
\usepackage{amsmath}
\usepackage{graphicx,tabularx,epsfig}
\usepackage{comment}
\usepackage{color}

\def\beq{\begin{equation}}
\def\eeq{\end{equation}}
\def\be{\begin{equation}}
\def\ee{\end{equation}} 
\def\bea{\arraycolsep .1em \begin{eqnarray}}
\def\eea{\end{eqnarray}}

\def\s0#1#2{\mbox{\small{$ \frac{#1}{#2} $}}}
\def\0#1#2{\frac{#1}{#2}}

\begin{document}
\graphicspath{{FIGURE/}} 
 
\title{Stability  
and UV completion of the Standard Model}

\author{Vincenzo Branchina and Emanuele Messina}
\affiliation{Department of Physics, University of Catania 
and INFN, Sezione di Catania, Via Santa Sofia 64, I-95123 
Catania, Italy }  

\begin{abstract}

The knowledge of the stability condition of the electroweak 
(EW) vacuum is of the greatest importance for our understanding 
of beyond Standard Model (BSM) physics. It is widely believed 
that new physics that lives at very high energy scales should have 
no impact on the stability analysis. This expectation has been 
recently challenged, but the results were controversial as  
new physics was given in terms of non-renormalizable 
higher order operators. Here we 
consider for the first time a renormalizable (toy) UV completion 
of the SM, and definitely show that such a decoupling does not 
take place. This result has important phenomenological 
consequences, providing a very useful test for BSM 
theories. In particular, it shows that speculations based 
on the so called ``criticality'' do not appear to be well founded.  
\end{abstract}

\maketitle
 
{\it Introduction.---} 
The analysis of the EW vacuum stability condition is of the greatest 
importance for our understanding of beyond  Standard 
Model (BSM) physics. Due to the top loop corrections, 
the Higgs potential $V(\phi)$ turns over for values of 
$\phi > v $, where $v \sim 246$ GeV is the location of the  
EW minimum, and develops a second minimum at a very large value 
$\phi_{min}^{(2)}$. The potential $V(\phi)$ is obtained by considering SM 
interactions only\,\cite{cab, sher, jones, sher2, alta, quiro, shapo, isido, isiuno, isidue,bu}, and depending on the Higgs and top masses, 
$M_H$ and $M_t$, the second minimum can be  higher or lower than 
the EW one. When $V(\phi_{min}^{(2)}) < V(v)$, the EW minimum is 
a metastable state ({\it false vacuum}), and we have to consider 
its lifetime $\tau$.
Fig.\,1 shows the usual stability phase diagram in the $M_H-M_t$ plane. 
The  stability, instability, and metastability regions 
are respectively for: $V(v) < V(\phi_{min}^{(2)})$; 
$V(\phi_{min}^{(2)}) < V(v)$ and $\tau < T_U$ ($T_U =$ age of 
the Universe); $V(\phi_{min}^{(2)}) < V(v)$ and 
$\tau > T_U$.  

When $V(\phi_{min}^{(2)}) < V(v)$, the {\it instability scale}\, 
$\phi_{inst}$\, of the Higgs potential is the value of the field where 
$V(\phi_{inst})=V(v)$, and $V(\phi) < V(v)$ for $\phi > \phi_{inst}$. 
For the present central experimental values of the Higgs and 
top masses, $M_H \sim 125.09$ GeV 
and $M_t \sim 173.34$ GeV \,\cite{higgsmass,ATLAS:2014wva}, it turns out that 
$\phi_{inst} \sim 10^{11} {\rm GeV} >> v$, 
$\phi_{min}^{(2)}\sim 10^{30}$ GeV, and  $\tau$ is much larger 
than $T_U$. 
Naturally, new physics interactions are expected to have 
an effect long before the scale $\phi_{min}^{(2)}\sim 10^{30}$ 
GeV is reached. The analysis outlined above is done under the 
assumption that new physics shows up only at very high energy 
scales, possibly the Planck scale. Moreover, 
it is assumed that, despite the presence of these new physics 
interactions, $\tau$ can be calculated 
with the potential obtained with SM interactions 
only\,\cite{isido, Espinosa:2007qp}. In fact, it is argued   
that the relevant scale for tunneling is the instability scale 
$\phi_{inst} \sim 10^{11}$ GeV, and that the contribution 
to the tunneling rate coming from very high scale physics 
($ >> \phi_{inst}$)
should be suppressed (decoupling)\,\cite{Espinosa:2007qp}. 
 
Contrary to these expectations, there are 
indications\,\cite{our1,our2,our3} that the presence of new physics 
at high energy scales can strongly modify the stability analysis. 
In refs.\,\cite{our1,our2,our3}, however, new physics interactions are 
parametrized in terms of few higher order, {\it non-renormalizable} 
operators. For this reason, these results are considered 
with a certain skepticism, and it is suggested that when the infinite 
tower of higher dimensional (new physics) operators of the 
{\it renormalizable} UV completion of the SM is taken into 
account, the effect should disappear, and the expected decoupling
should be recovered. Actually, it is thought that this 
effect takes place above the physical cutoff, where 
the control of the theory is lost \cite{Espinosa:2015qea}.

In this Letter we consider for the first time 
a {\it fully renormalizable} (toy) 
UV completion of the SM, where new physics interactions live 
at scales {\it much higher than the instability scale} $\phi_{inst}$, 
and perform the stability analysis of the EW vacuum.
We shall then be able to provide a definite 
answer to the crucial question of whether the stability condition 
of the EW vacuum is affected by the presence of very high scale 
new physics or, as commonly expected,
a decoupling takes place\,\cite{Espinosa:2007qp}. 
A clear understanding of this issue is of the greatest importance 
for BSM physics, provides very useful guidance for BSM
model building, and is the main motivation for the present work. 

{\it The model.---} The classical potential for the Higgs doublet 
\begin{eqnarray}\label{Higgsdoub}
\Phi=\frac{1}{\sqrt{2}}\left(\begin{array}{c}
-i(G_1-iG_2)\\
\phi + i G_3
\end{array}\right)\,,
\end{eqnarray}
where $\phi$ is the Higgs field and $G_i$ are the Goldstones, is:
\begin{eqnarray}\label{pot}
U(\Phi)=m^2 \left(\Phi^\dagger\cdot \Phi\right)+\lambda
\left(\Phi^\dagger \cdot \Phi\right)^2\,.
\end{eqnarray}

Our (toy) renormalizable UV completion of the SM is 
obtained by considering the addition of a scalar field $S$
and a fermion field $\psi$ that interact in a simple way 
with $\Phi$, and have masses $M_S$ and $M_f$ well above 
the instability scale, $M_S , M_f >> \phi_{inst}$. 
Apart from the kinetic terms, 
the additional terms in the Lagrangian are: 
\begin{eqnarray}\label{v1}
\Delta \,{\cal L}&=&\frac{M_S^2}{2}S^2 + \frac{\lambda_S}{4} S^4
+ 2 g_S \left(\Phi^\dagger \cdot \Phi\right) S^2\nonumber\\
&+& M_f 
\left(\bar\psi_L\psi_R+\bar\psi_R\psi_L\right)\nonumber\\
&+&\sqrt{2} g_f \left(\bar \Psi_L\cdot \Phi \, \psi_R 
+\bar \psi_R \,\Phi^\dagger \cdot  \Psi_L \right)\,,
\end{eqnarray}
where $\lambda_S$ is the self-coupling of the 
new scalar $S$, $g_S$ the coupling between the 
Higgs doublet and $S$, 
$\psi_L$ and $\psi_R$ the left and right components 
($SU(2)$ singlets) of the Dirac field $\psi$ with mass $M_f$,  
$\Psi_L$ the left-handed $SU(2)$ fermion   
doublet $\Psi_L=(0,\psi_L)^T$ (we are not considering  
additional neutrinos), and $g_f$ the Yukawa 
coupling between $\Psi$ and the Higgs doublet
$\Phi$. 

\begin{figure}[t]
\vskip-4mm
\includegraphics[width=.44\textwidth]{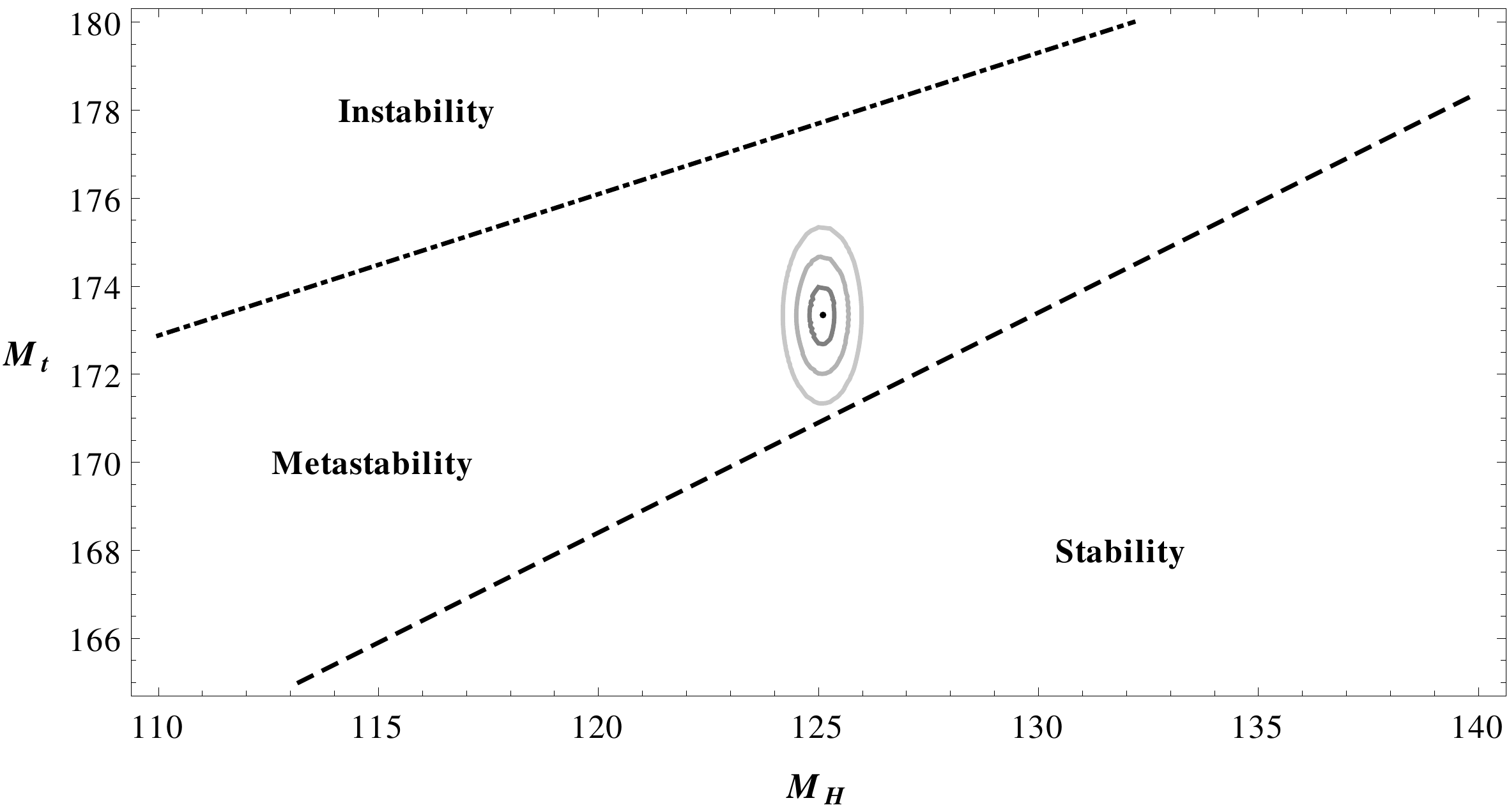}\vskip-4mm
\caption{
This figure shows the usual stability diagram for the EW vacuum,  
done under the assumption that new physics interactions at very 
high scales have no impact on its stability 
condition\,\cite{isido, isiuno, isidue}. The 
$M_H-M_t$ plane is divided in three sectors (see text): absolute 
stability, metastability and instability regions. 
The dot indicates the current central experimental values 
$M_H\sim 125.09$ GeV and $M_t\sim 173.34$ GeV, and the 
corresponding EW vacuum lifetime is 
$\tau \sim 10^{600} \, T_U$ (see text).
}
\label{bounn}\vskip-.6cm
\end{figure}

Inserting (\ref{Higgsdoub}) in (\ref{v1}) we have:
\begin{eqnarray}\label{int}
\Delta \,{\cal L}&=&\frac{M_S^2}{2}S^2 
+ \frac{\lambda_S}{4} S^4 + g_S \varphi^2 S^2 
+M_f \bar\psi\psi\nonumber\\
&+&g_f \varphi \bar \psi \psi
+g_S \left(G_1^2+G_2^2+G_3^2 
\right)S^2\nonumber\\
 &+&g_f G_3 \bar\psi\left[\left(\frac{1+\gamma_5}{2}
\right)
+i\left(\frac{1-\gamma_5}{2}\right)\right]\psi 
. 
\end{eqnarray}  

For the purposes of the present work, it is sufficient to 
consider the impact of these additional terms on the Higgs potential  
$V(\phi)$ at the one-loop level only. Then, in the 
following we do not need to consider further the second and 
the third lines of Eq.\,(\ref{int}). The one-loop contribution 
to $V(\phi)$ from the 
additional terms in $\Delta \,{\cal L}$ is: 
\begin{eqnarray}\label{renormpot}
V_1(\phi)&=&
\frac{\left(M_S^2+2g_S\phi^2\right)^2}{64\pi^2}
\left[\ln\left(\frac{M_S^2+2g_S\phi^2}{M_S^2}\right)
-\frac{3}{2}\right]\nonumber\\
&-&\frac{\left(M_f^2+g_f^2\phi^2\right)^2}{16\pi^2}
\left[\ln\left(\frac{M_f^2+g_f^2\phi^2}{M_S^2}\right)
-\frac{3}{2}\right] \, 
\end{eqnarray}
where the renormalization scale is chosen at $\mu=M_S$.

According to the decoupling argument\,\cite{isido,Espinosa:2007qp}, 
these new physics interactions at very high energy scales 
($M_S, M_f >> \phi_{inst} \sim 10^{11}\,{\rm GeV}$) 
should have no impact on the  stability analysis. We now 
investigate this question by considering two 
choices for the parameters of our (toy) UV completion of the SM.  
In both cases, a second minimum deeper than the EW one is 
formed, and we then have to calculate the EW vacuum lifetime.

{\it Results.---} 
We now impose to the modified potential 
$V(\phi)=\frac{\lambda}{4}\,\phi^4 + V_1(\phi)$ (as usual the
quadratic term can be neglected as we consider very high 
values of $\phi$) the matching conditions at the threshold 
scale $M_f$ so that the SM Higgs potential is recovered for 
values of $\phi<M_f$. The EW vacuum lifetime  
$\tau$ is then given by\,\cite{our3}
\be \label{Tunn}
\tau= \min_{\mu} 
\left(\frac{1}{T_U^3\mu^4}
\exp{\frac{8\pi^2}{(3|\lambda_{SM}(\mu)+4
\overline V_1(\mu)/\mu^4 |)}}\right),
\ee
where $\lambda_{SM}(\mu)$ is the running quartic 
coupling, and $\overline V_1(\phi)$ is nothing but the 
additional contribution (\ref{renormpot}) to the Higgs 
potential with the $\phi^2$ and $\phi^4$ terms subtracted. 

Before going on with the calculation of $\tau$ for our model, 
it is worth to remind that when the presence of these new 
physics interactions (usually assumed to be harmless for 
calculating $\tau$) is neglected, for the present  
central values of the Higgs and top masses, 
$M_H=125.09$ GeV and $M_t=173.34$ 
GeV \,\cite{higgsmass,ATLAS:2014wva}, the calculation gives:  
\be\label{tuntime1}
\tau \sim 10^{600} \, T_U\,.  
\ee
 
This result is the basis for the so called metastability 
scenario, according to which although the EW minimum is a 
metastable state (and then a false vacuum), as its
lifetime turns out to be much larger than the age of 
the universe, we may well live in such a state. 

Fig.1 shows the full stability analysis done under the 
assumption that new physics at scales $>> \phi_{inst}$ has  
no impact on the stability condition of the EW 
vacuum\,\cite{isido, isiuno, isidue,bu}. 
The black dot corresponds to the tunneling time of 
Eq.\,(\ref{tuntime1}). The ellipses give the one, two and 
three sigma experimental uncertainties.

We move now to the computation of the EW vacuum lifetime for 
our model with new physics at very high 
energy scales ($>> \phi_{inst}$).
For our first example, we choose $M_S = 1.2\,\cdot 10^{18}$ GeV, 
$M_f =0.6 \,\cdot 10^{17}$ GeV, $g_S(M_S) =0.97$, 
$g_f^2 (M_S)=0.48$, $\lambda_{SM} (M_S)=-0.0151$. The latter
is the value of the running quartic coupling $\lambda_{SM} (\mu)$
at the scale $\mu=M_S$, obtained  by considering the RG 
equations for the SM coupling constants and the boundary 
conditions at the next-to-next 
to leading order\,\cite{isidue, NNLO}. 
Note that at the order of approximation
that we are considering, $\lambda_S$ plays no role.

For the values of the parameters given above, the Higgs effective 
potential $V(\phi)$ develops a new minimum, lower 
than the EW one, at $\phi_{min}\sim 0.4 \cdot 10^{19}$\, GeV. 
To study the stability condition 
of the EW vacuum, we have then to calculate its 
lifetime $\tau$.
For the present central  experimental 
values of the Higgs and top masses ($M_H=125.09$ GeV and 
$M_t=173.34$ GeV) we find: 
\be\label{tuntime2}
\tau \sim 10^{180} \, T_U\,.  
\ee
 
This result has to be compared with the tunneling time 
of Eq.\,(\ref{tuntime1}), obtained by considering 
the SM potential alone (no new physics  
included). Although for this example the tunneling time is still
much higher than the age of the Universe, Eq.\,(\ref{tuntime2})
gives a result that is greatly different from the one of  
Eq.(\ref{tuntime1}).

Let us consider now another example, with a different value of 
$M_f$. More precisely, let us take 
$M_S = 1.2\,\cdot 10^{18}$ GeV, 
$M_f =2.4 \,\cdot 10^{15}$ GeV, 
$g_S (M_S) =0.97$, $g_f^2 (M_S) =0.48$, and 
$\lambda_{SM} (M_S)=-0.0151$. Again for the present central  
experimental values of $M_H$ and $M_t$ we find: 
\be\label{tuntime3}
\tau \sim 10^{-65} \, T_U\,.  
\ee

In this case, the situation is more dramatic than in the 
previous example: the tunneling time turns out to be much smaller
than the age of the Universe. If realistic, this model could not 
be accepted.

The lesson from Eqs.\,(\ref{tuntime1}), (\ref{tuntime2}), and 
(\ref{tuntime3}) is clear. The 
expectation that the tunneling time should be insensitive 
to physics that lives at energies higher than the instability
scale, in other words that the result shown in Eq.\,(\ref{tuntime1})
should not be modified by the presence of new physics at high 
energies, is not fulfilled.

\begin{figure}[t]
\vskip-4mm
\includegraphics[width=.44\textwidth]{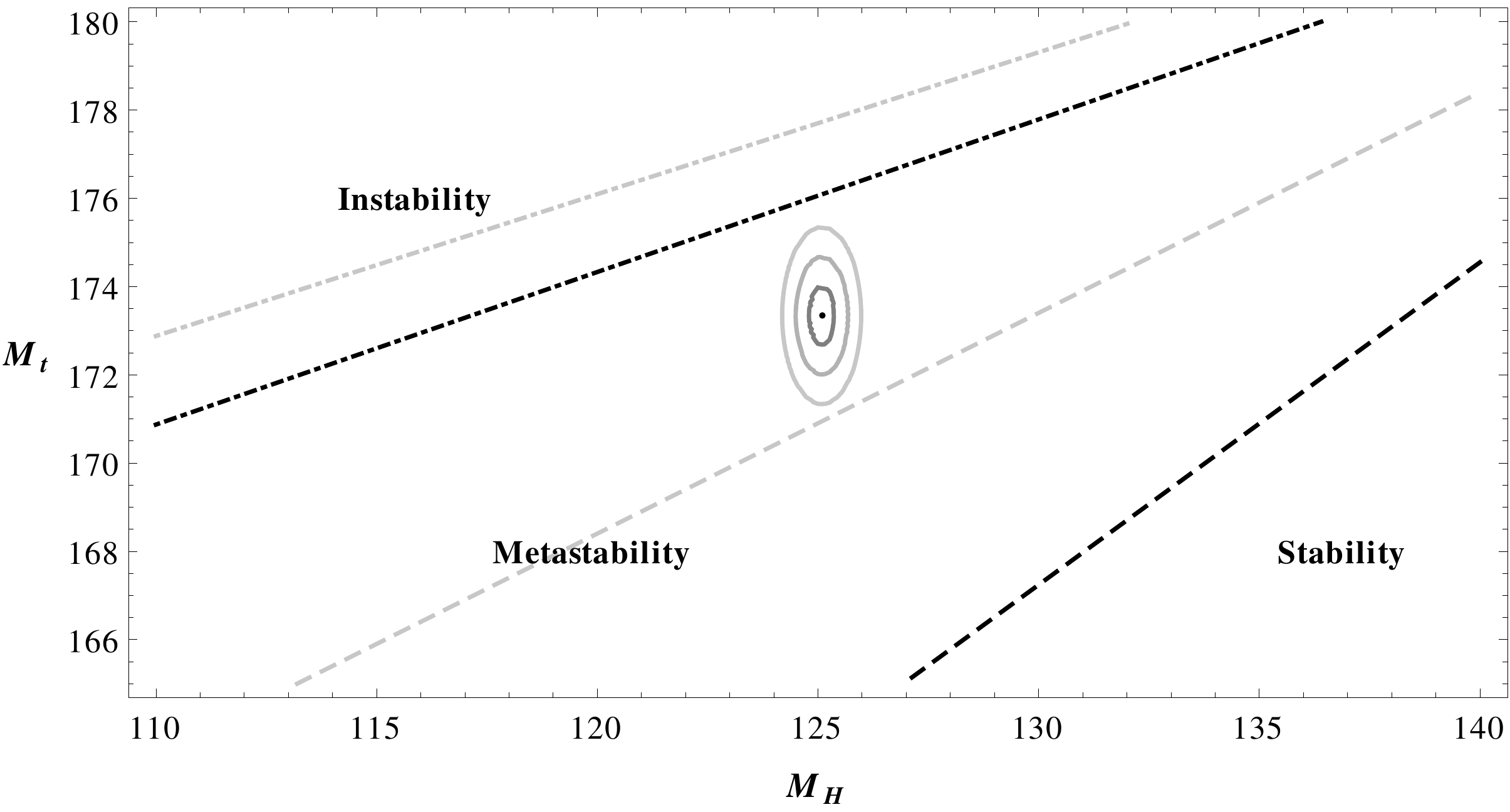}\vskip-4mm
\caption{
This figure shows the stability diagram for the EW vacuum in the 
$M_H-M_t$ plane when the toy (very high energy) UV completion of 
the SM given in 
Eq.\,(\ref{v1}) is considered. The Higgs effective potential is modified 
by the presence of the additional term (\ref{renormpot}). 
Here $M_S = 1.2\,\cdot 10^{18}$ GeV, 
$M_f =0.6 \,\cdot 10^{17}$ GeV, 
$g_S =0.97$, $g_f^2 =0.48$. 
As in fig.1, the $M_H-M_t$ plane is divided in three sectors: 
absolute stability, metastability and instability regions. 
The presence of the term (\ref{renormpot}) causes the
lowering of the instability and stability lines. 
}
\label{bounn}\vskip-.6cm
\end{figure}

Naturally we may ask why the decoupling 
argument fails. The reason is that the decoupling 
theorem applies when we calculate scattering amplitudes at 
energies $E$  lower than $M_S$ and $M_f$. 
In these cases, the contributions from  high energy new 
physics is suppressed by factors 
as $E/M_S$ and $E/M_f$ to some appropriate power.

In our case, however, we are calculating the tunneling time. 
Tunneling is a non-perturbative phenomenon, and no 
decoupling applies: in the 
calculation of $\tau$, no naive suppression factor, 
$\phi_{inst}/M_S$ or $\phi_{inst}/M_f$, appears.  
More technically, the tunneling time 
$\tau$ is essentially given by the exponential in Eq.\,(\ref{Tunn}), 
whose argument is the action calculated at bounce 
solution to the euclidean Euler-Lagrange equation of 
motion \cite{coleman}.  
If the Higgs potential is modified by the presence of terms 
as the one in Eq.\,(\ref{renormpot}), the new bounce turns out to be
different from the one obtained when this term is
absent. The action is  
modified and (once exponentiated) gives rise to a 
value for $\tau$ that can be enormously different  
from the result obtained when new physics is not 
considered. 

This is a central result of the present work. 
With the help of a fully renormalizable (toy) UV completion 
of the SM, we have firmly shown that, contrary to a widely 
diffused expectation, the EW vacuum lifetime strongly
depends on new physics even if the latter lives at very high 
energy scales, much higher than the instability scale 
$\phi_{inst} \sim 10^{11}$ GeV. As we have just shown, this 
phenomenon is not due to an illegitimate extrapolation of 
the theory beyond its validity\,\cite{Espinosa:2015qea}. 

On the contrary, it is an illegitimate application of the 
decoupling argument to a phenomenon to which it cannot be applied, 
namely the (non-perturbative) tunneling phenomenon, that 
leads to the expectation that physics 
at scales much higher than the instability 
scale $\phi_{inst}$ should have no impact on the stability 
condition. 

We are now ready to proceed with our analysis. Fig.1  
shows the stability diagram in the $M_H-M_t$ plane  
obtained under the assumption (decoupling argument)
that the stability analysis does not depend on high 
energy physics. 
The examples considered above, with the results 
(\ref{tuntime1}), (\ref{tuntime2}) and (\ref{tuntime3}), 
indicate that we should expect that the whole stability 
phase diagram actually depends on new physics, even if 
it lives at very high energy scales.   

The dashed and 
the dashed-dotted lines of fig.\,1 are named the 
stability line and the instability line respectively. 
The first one
is obtained for those values of $M_H$ and 
$M_t$ such that the two minima are at the same height,
the latter is obtained for the case when 
$V(\phi^{(2)}_{min}) < V(v)$ and $\tau=T_U$.  

Let us repeat now the stability analysis when the 
term (\ref{v1}), that is our (toy) UV completion 
of the SM, is added to the SM Lagrangian, so that the term 
(\ref{renormpot}) is added to the Higgs effective potential. 
In fig.\,2, the analysis is performed for the 
values of the parameters considered in our first example, namely 
$M_S = 1.2\,\cdot 10^{18}$ GeV, 
$M_f =0.6 \,\cdot 10^{17}$ GeV, 
$g_S (M_S) =0.97$, $g_f^2(M_S) =0.48$, 
$\lambda_{SM} (M_S)=-0.0151$. 

We note that the instability line moves downwards. This result 
had to be expected from the previous results (\ref{tuntime1}) 
and (\ref{tuntime2}) for the tunneling time.  
In fact, we obtained $\tau \sim 10^{180}\, T_U$ for the UV completed 
Higgs potential and $\tau \sim 10^{600}\, T_U$ for the SM Higgs 
potential. It is clear that in the case of the UV completed 
potential, the experimental point (black dot) must be closer to the 
instability line than in the case of the unmodified potential. 
The grey lines of fig.\,2 are the old instability and 
stability lines for the the unmodified Higgs potential 
(see fig.\,1).

Actually, another important effect is that even the stability 
line moves downwards 
(see fig.\,2). When it was thought that a decoupling 
effect assured that new physics at high scales could not modify 
this diagram,  many speculations were triggered by the fact that 
the experimental point (black dot in the figure)  
$M_H \sim 125.09$ GeV and  $M_t\sim 173.34$ GeV lies 
``close'', within 2-3 sigma, to the stability line.

\begin{figure}[t]
\vskip-4mm
\includegraphics[width=.44\textwidth]{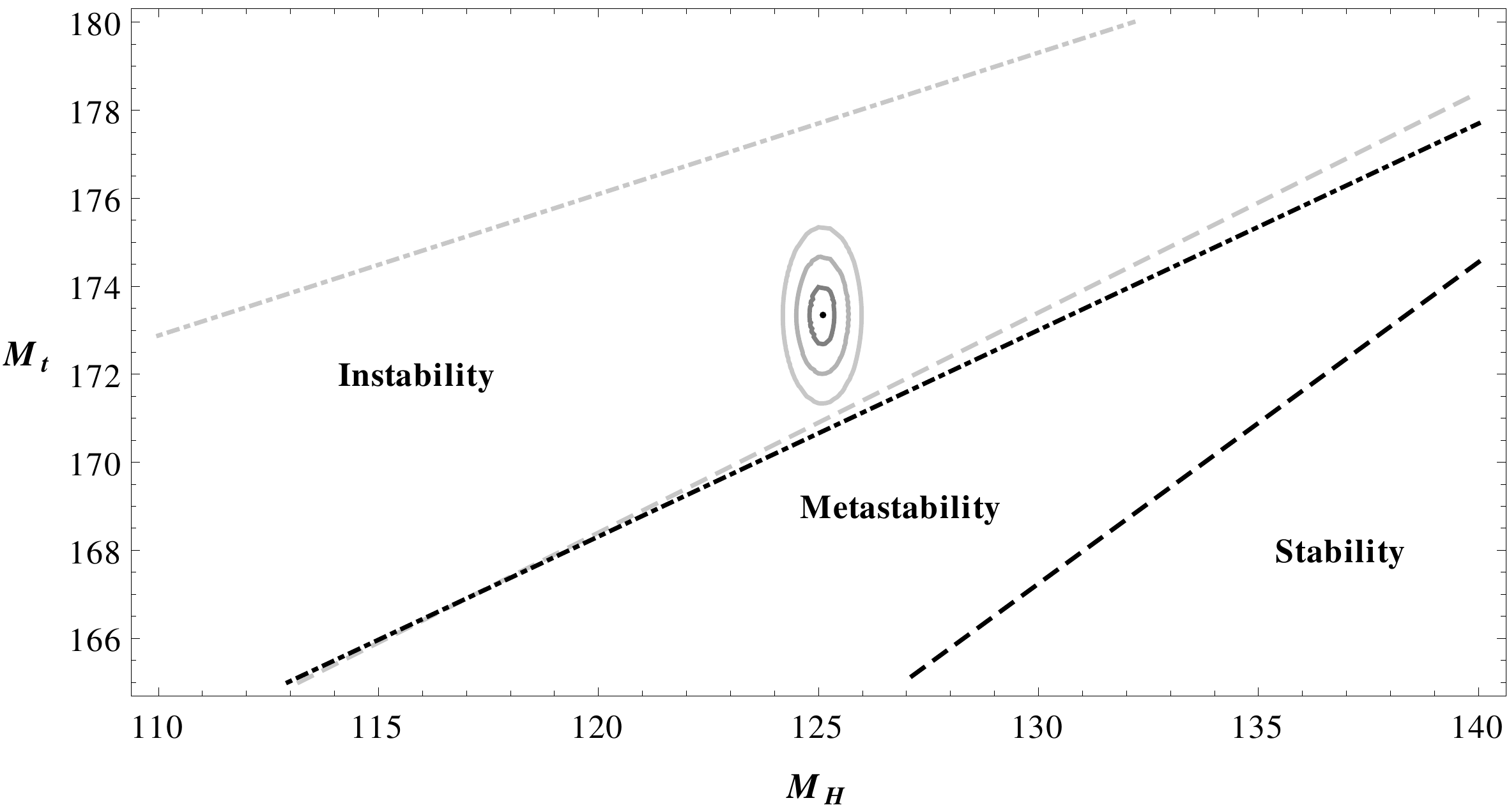}\vskip-4mm
\caption{
The same as in fig.\,2, for different values of the parameters.
Here $M_S = 1.2\,\cdot 10^{18}$ GeV, 
$M_f =2.4 \,\cdot 10^{15}$ GeV, $g_S =0.97$, $g_f^2 =0.48$. 
As in fig.2, the presence of the term (\ref{renormpot}) causes 
the lowering of the instability and the stability lines. 
However, in this case the instability line goes below the 
experimental point, signaling that the model, for these values
of the parameters, cannot be considered as a viable UV completion 
of the SM.
}
\label{bounn}\vskip-.6cm
\end{figure}

In this respect, it was suggested that more refined measurements 
of $M_t$ and $M_H$ should allow to determine whether the 
EW vacuum is a stable or a metastable state. Some 
authors even went to the point to consider the  
closeness of the experimental point to the stability line
as the most important message from LHC\,\cite{bu}, speculating 
on this closeness and elaborating on it for 
model building\,\cite{Espinosa:2015qea}. 

The results presented in this letter show that 
the stability condition of the EW vacuum is much more sensitive 
to high energy new physics than to the values of the Higgs 
and top masses. Therefore, more refined measurements of $M_t$ 
and $M_H$, that are clearly very important on their own, 
will not allow to determine the stability condition of the 
EW vacuum. 

Moreover, speculations and model building inspired by the 
so called ``criticality condition'',   
the closeness of the experimental point to the stability 
(also called critical) line, are actually unjustified. 
As we have seen, new physics even if it lives at very high 
energies (we certainly expect new physics at least at very
high energies, maybe Planck scale) can enormously 
modify the distance between the experimental point and 
the critical line.   

Finally, in fig.\,3 the stability diagram for our model 
with the values of the parameters considered in our 
second example ($M_S = 1.2\,\cdot 10^{18}$ GeV, 
$M_f =2.4 \,\cdot 10^{15}$ GeV, 
$g_S (M_S)=0.97$, $g_f^2 (M_S)=0.48$, 
$\lambda_{SM} (M_S)=-0.0151$)
is shown. The instability and stability lines
move downwards as for the previous case. 
In this case, however, the tunneling time for the experimental
point is much shorter than the age of the Universe, see 
Eq.\,(\ref{tuntime3}), and in fact we see that the experimental 
point is now inside the instability region. 

This simply means that the model with these values of the 
parameters cannot be considered as a viable UV completion of 
the SM. This result contains another important lesson of the 
present work. The stability condition of 
the EW vacuum, as we have shown, is strongly sensitive to 
new physics, even when it lives at very high energy scales. 
Therefore, as we have shown that for the stability analysis 
we cannot rely on the 
decoupling of high energy physics, we conclude that 
candidate BSM theories have to be  
checked against a sort of stability test. Only 
models with a stable or metastable (but with $\tau > T_U$ ) 
EW vacuum can be considered as viable UV completions of 
the SM. 

{\it Conclusions.---} With the help of a (toy) renormalizable 
UV completion of the SM, we have definitely shown that new 
physics interactions, even when they live at energies 
much higher than the scale where the Higgs potential 
becomes unstable (the so called instability scale 
$\phi_{inst} \sim 10^{11}$ GeV),
have strong impact on the stability condition of the EW 
vacuum. With respect to previous analyses, here new physics
interactions are given in terms of a fully renormalizable theory 
rather than with the help of higher order non-renormalizable 
operators, and this makes the conclusions of the present work 
really robust. These results have far reaching phenomenological
consequences, providing very useful guidance for BSM model 
building. In particular, they show that speculations based 
on the so called ``criticality'' are not well founded.

\end{document}